\documentclass[preprint,showpacs,floatfix]{revtex4-1}
\usepackage{float}
\usepackage[caption = false]{subfig}
\usepackage[final]{graphicx}

\usepackage{dcolumn}% Align table columns on decimal point
\usepackage{bm}% bold math
\usepackage{amssymb,amsmath}
\usepackage{color}
\usepackage{epstopdf}

\begin{document}
\title{Micro-plasticity in a fragile model binary glass}
\author{P. M. Derlet}
\email{Peter.Derlet@psi.ch} 
\affiliation{Condensed Matter Theory Group, Paul Scherrer Institut, CH-5232 Villigen PSI, Switzerland}
\author{R. Maa{\ss}}
\affiliation{Department of Materials Science and Engineering, University of Illinois at Urbana Champaign, 1304 West Green Street, Urbana, Illinois 61801, USA}
\affiliation{Federal Institute of Materials Research and Testing (BAM), Unter den Eichen 87, 12205 Berlin, Germany}	
\date{\today}
	
\begin{abstract}
Atomistic deformation simulations in the nominally elastic regime are performed for a model binary glass with strain rates as low as $10^{4}$/sec (corresponding to 0.01 shear strain per 1$\mu$sec). A robust elasticity is revealed that exhibits only minor elastic softening, despite quite different degrees of structural relaxation occurring over the four orders of magnitude strain rates considered. A closer inspection of the atomic-scale structure indicates the material response is distinctly different for two types of local atomic environments. A system spanning iscosahedrally coordinated substructure responds purely elastically, whereas the remaining substructure also admits microplastic evolution. This leads to a heterogeneous internal stress distribution which, upon unloading, results in negative creep and complete residual-strain recovery. A detailed structural analysis in terms of local stress, atomic displacement, and SU(2) local bonding topology shows such microscopic processes can result in large changes in local stress and are more likely to occur in geometrically frustrated regions characterized by higher free volume and softer elastic stiffness. These insights shed atomistic light onto the structural origins that may govern recent experimental observations of significant structural evolution in response to elastic loading protocols.
\end{abstract}
\maketitle

\section{Introduction}

The amorphous solid is a material strongly out of equilibrium and very much dependent on its thermal and loading history~\cite{Sun2016,Hufnagel2016}. Despite this, structural glasses have a robust and extended elastic regime~\cite{Tian2012}, and a well defined and high yield stress~\cite{Wu2008,Harmon2007}, giving them real-world industrial and engineering applications. This robustness and reproducibility implies a strongly constrained microstructure and therefore a spatially-correlated structural disorder from which structural length-scales can emerge~\cite{Zhu2017,Liu2018}. Such emergent structural heterogeneity, which is difficult to characterize experimentally, strongly influences the macroscopic plasticity of the amorphous solid~\cite{Wang2018,An2016,Zhu2016}.

Classical thermally-activated micro-plasticity during the nominally elastic region, offers direct insight into a class of fundamental excitations of the solid that exist also under zero-load conditions~\cite{Maass2018}. Signatures of such microplastic activity can, for example, be indirectly seen in micro-tensile experiments~\cite{Wu2009}, creep and damping~\cite{Castellero2008,Ye2010}, isostatic macroscopic loading~\cite{Lee2008,Greer2016,Lei2019}, fatigue~\cite{Ross2017}, or stress-relaxation experiments~\cite{Ross2017}. The attribute ``indirect'' is used here since deformation in the absence of resolvable shear-band formation, is resolved as either non-linear pre-yield behavior, a permanent residual strain upon unloading, or an increase in the stored excess enthalpy. More specific examples are the rejuvenation and a local increase in nano-scale elastic heterogeneity after cyclic loading~\cite{Ross2017}, or the residual creep strain after applying a constant load at low homologous temperature~\cite{Lee2008,Greer2016}. Such studies reveal changes in measurable quantities that imply structural changes which, within the precision of experiments, can hardly be resolved. This inability to track detailed structural changes can be somewhat overcome with novel scattering methods using both x-rays and/or electrons, where for example fluctuation microscopy inside a transition electron microscope~\cite{Zhang2018} or speckle patterns from coherent x-ray scattering~\cite{Ruta2017} can shed light on the atomic scale structural change or dynamics. The present authors pursued such an approach in very recent work by tracking the structural dynamics during nominally elastic loading~\cite{Das2019} and during cryogenic protocols~\cite{Das2020} that activate structural excitations underlying the Gamma-mode~\cite{Kuechemann2017}. Both these efforts reveal a rich and non-trival signature of structural dynamics quantified with time-dependent relaxation times that hint towards thermally-activated microplastic processes heterogeneous in both space and time. However, the quantified signals remain an average of a large number of atoms, and a detailed picture of the spatio-temporal characteristics of individual structural excitations is lacking.

This is where atomistic simulations have for a long time played a pivotal role by providing model based understanding of the microscopic structure and processes of metallic glasses, with a primary focus on binary and ternary model glass systems. Such efforts strongly support the early ideas of Frank~\cite{Frank1952}, Chaudhari and Turnbull~\cite{Chaudhari1978} and Nelson~\cite{Nelson1983a,Nelson1983b}, which asserted a short-range order dominated by 5-fold coordinated bonds characterized by localized structural motives with (or close to) an icosahedral symmetry. Indeed, the fraction of local icosahedral environments is a strong measure of the degree to which a glass is relaxed~\cite{Sheng2006,Ding2014,Ding2017,Derlet2017a,Derlet2018,Derlet2020,Derlet2020a}, and the constraints associated with their connectivity is one avenue to understand medium range order and the associated emergence of structural length-scales~\cite{Derlet2020a,Ma2015}. Simulation has also given fundamental insight into the underlying microscopic processes during stress-driven athermal plasticity, demonstrating that mechanisms which mediate the collective athermal macroscopic response are fundamentally local~\cite{Falk1998,Maloney2004,Demkowicz2005}. Such localized activity is compatible with early thermal plasticity theories~\cite{Spaepen1977,Argon1979} and their coarse-grained simulation implementations~\cite{Bulatov1994a,Bulatov1994b,Bulatov1994c,Homer2009,Homer2010,Li2013}. More recently, potential energy landscape exploration methods also revealed a local picture for thermally-activated structural excitations under both zero~\cite{RodneyPRL2009,RodneyPRB2009,Koziatek2013,Swayamjyoti2014,Fan2017} and applied load~\cite{Swayamjyoti2016} --- a result confirmed by finite temperature molecular dynamics  simulations~\cite{Derlet2017,Derlet2017a,Derlet2018} suggesting the possibility of glass plasticity driven by thermal fluctuations at sufficiently long deformation time scales.

Here we employ finite temperature molecular dynamics at the micro-second timescale, to demonstrate that a well relaxed binary model glass system exhibits a robust elastic regime largely insensitive to any structural relaxation occurring during the elastic loading. We do this for a well known Lennard-Jones system~\cite{Wahnstrom1991}, whose resulting glassy structure forms a percolative icosahedral network consisting of fragments of the C15 Laves phase~\cite{Derlet2020a} --- a feature that material specific model systems also exhibit~\cite{Zemp2014,Zemp2016}. We consider deformation strain rates spanning four orders of magnitude, and find that apart from minor elastic softening, the elastic regime is largely insensitive to the considered timescales, even though at the lowest strain rate of $10^{4}$/sec, significant thermally-driven structural relaxation and micro-plasticity occurs. Central to this insensitivity is a pre-existing percolative icosahedral network that elastically deforms, whilst the remainder of the material undergoes micro-plasticity, resulting in a heterogeneous material in terms of internal stress. Unloading results in a residual strain, with the resulting elastic back-stress of the icosahedral network causing thermally-driven negative creep that leads to full recovery of the residual strain at a timescale comparable to the initial deformation.  

\section{Simulation methodology}

\subsection{Choice of Lennard-Jones potential, 50:50 chemical composition and physical units}

The binary Lennard-Jones potential of Wahnstr\"{o}m~\cite{Wahnstrom1991} has been used extensively in atomistic simulations of model binary glass systems~\cite{Wahnstrom1991,Pedersen2010,Rodney2011,Koziatek2013,Swayamjyoti2014,Swayamjyoti2016,Derlet2017,Derlet2017a,Derlet2018,Derlet2020} and is widely known as a model fragile glass former. Since it is a pair potential, it is unable to describe the unsaturated nature of the metallic bond, and therefore the quantitative aspects of material specific binary model alloys. For this, embedded atom~\cite{Daw1984} or second-moment~\cite{Finnis1984} empirical many-body potentials are needed. Despite this aspect, the potential captures the essential structural physics of bulk binary metallic glasses as for example CuZr, ZrNi, NiNb, HfCu, CaAl etc., and the competition between liquid-like and crystal-like bonds which is believed to underlie such metallic glasses and its medium range order~\cite{Frank1952,Chaudhari1978,Sheng2006,Ma2015}. Fundamentally this is due to the scale of local density variations, which whilst important for characterizing the structural state of the amorphous solid, are too small to distinguish the explicit many-body aspect of the material specific potential. For systems in which there are large changes in coordination, such as that at a surface or an internal cavitation, such Lennard Jones potentials would not be able to capture the bond stiffening expected for a metal. These aspects have been discussed in more detail in Refs.~\cite{Derlet2017,Derlet2017a,Derlet2018,Derlet2020}.

The present work employs a 50:50 chemical composition of small to large atoms, which is the potential's eutectic composition, therefore producing a homogeneous undercooled liquid and a glass with little chemical segregation.

The Lennard-Jones potential has the following form:
\begin{equation}
V_{ab}(r)=4\varepsilon\left(\left(\frac{r}{\sigma_{ab}}\right)^{12}-\left(\frac{r}{\sigma_{ab}}\right)^{6}\right)
\end{equation}
where $\sigma_{22}=5/6\sigma_{11}$ and  $\sigma_{12}=\sigma_{21}=11/12\sigma_{11}$ for the Wahnstr\"{o}m parametization. The atoms of type 1 may be considered as the larger atom type. The atomic masses of the two atom types are arbitrarily chosen such that $m_{1}/m_{2}=2$. For a molecular dynamics iteration, a time step of 0.002778$\sigma_{11}\sqrt{m_{1}/\varepsilon}$ is used. The distance unit is taken as $\sigma=\sigma_{11}$ and the energy unit as $\varepsilon$, with stresses in the units of $\varepsilon/\sigma^{3}$. For this work, the potential is truncated to a distance 2.5$\sigma$.

When metallic units (representative of say, CuZr) are taken for $\varepsilon$ and $\sigma$, an MD time-step is of the order of a femto-second. Thus one
billion MD iterations corresponds to approximately one micro-second. Throughout this paper, simulation times will be measured with respect to one billion MD iterations, which in turn is approximated as one micro-second. Absolute
temperatures are expressed as an energy, $k_{\mathrm{B}}T$, using a value of $k_{\mathrm{B}}=8.617\times10^{-5}\varepsilon$ where $\varepsilon$ is assumed to have units of electron-Volt.

\subsection{Sample preparation and deformation simulations} \label{sssample}

The sample preparation protocol exploits an initial NVT ensemble to produce a glassy structure at a local minimum in the PEL~\cite{Swayamjyoti2014,Swayamjyoti2016,Derlet2017a,Derlet2018,Derlet2020}. This sample has undergone 2 micro-seconds of relaxation at $0.95T^{\mathrm{NVT}}_{\mathrm{f}}$. Here, $T^{\mathrm{NVT}}_{\mathrm{f}}$ is the fictive glass transition temperature at which the quenched undercooled liquid first becomes a glass (see Ref.~\cite{Derlet2017}).  Fixed zero pressure NPT simulations are subsequently performed using a linear temperature annealing protocol to determine the new fictive glass transition temperature $T^{\mathrm{NPT}}_{\mathrm{f}}$. Here such NPT simulations allow for the volume to fluctuate isotropically. Earlier work demonstrated that the same structural state may be obtained using either NVT or NPT ensembles, or a combination of both~\cite{Derlet2018}. The configuration at $0.8T^{\mathrm{NPT}}_{\mathrm{f}}$ is then used for the shear deformation and zero-load simulations. In the present work, all simulations are performed at a temperature of $0.8T^{\mathrm{NPT}}_{\mathrm{f}}$. 

The NPT simple-shear deformation simulations are performed involving a fixed strain rate in the $xy$ plane. Thus the system's volume may fluctuate isotropically as the $xy$ strain is applied. Due to only isotopic volume fluctuations being allowed, and also the finite size of the sample, the final sample will contain a small non-zero global shear stress corresponding to small shear strains of order 0.001. Over several micro-seconds these components may fluctuate in both sign and magnitude. Compared to thermal stress fluctuations and the typical strains encountered in the present work they may be considered negligible. Their presence my be seen in the unloading curves of Fig.~\ref{figUnloading}, the creep recovery simulations in Fig.~\ref{figCreep} and the local stress analysis in Fig.~\ref{figLocalStress}.

\section{Deformation simulations}

\subsection{Loading/unloading} \label{ssecDeformation}

\begin{figure}
\begin{center}   
\subfloat[]{\includegraphics[width=0.55\linewidth,trim=1cm 0.5cm 1cm 1.5cm]{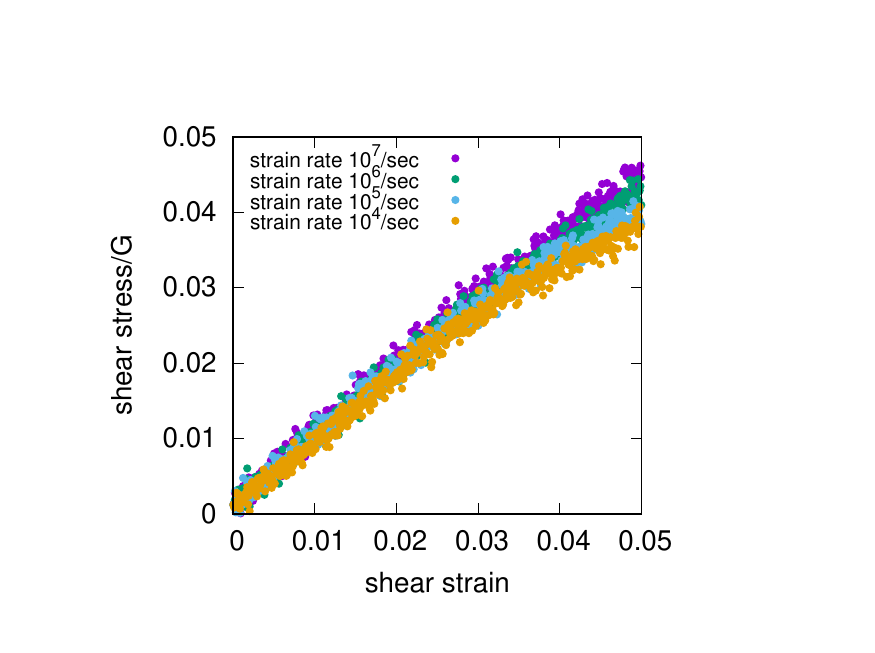}}
\subfloat[]{\includegraphics[width=0.55\linewidth,trim=1cm 0.5cm 1cm 1.5cm]{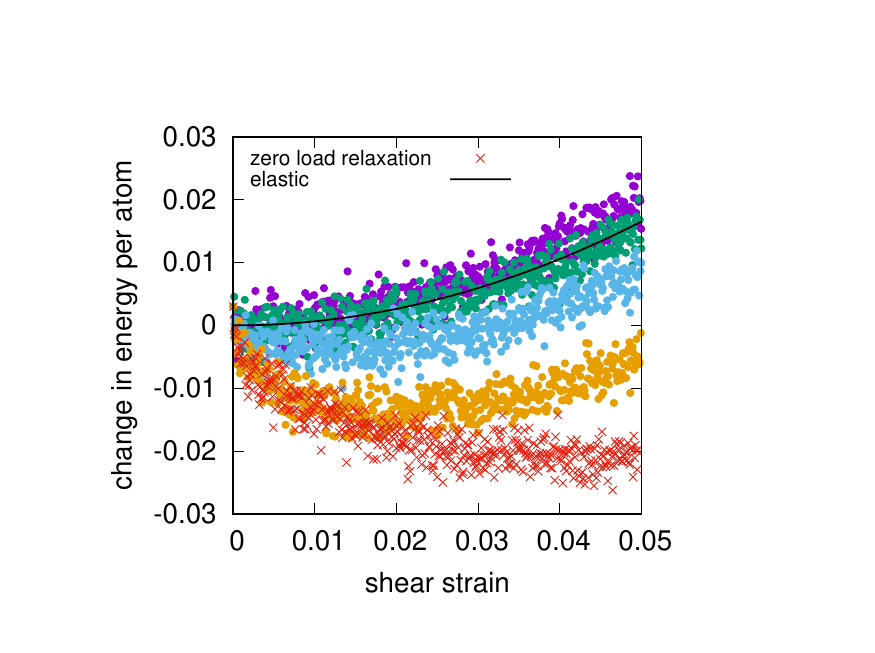}}\\
\subfloat[]{\includegraphics[width=0.55\linewidth,trim=1cm 0.5cm 1cm 1.5cm]{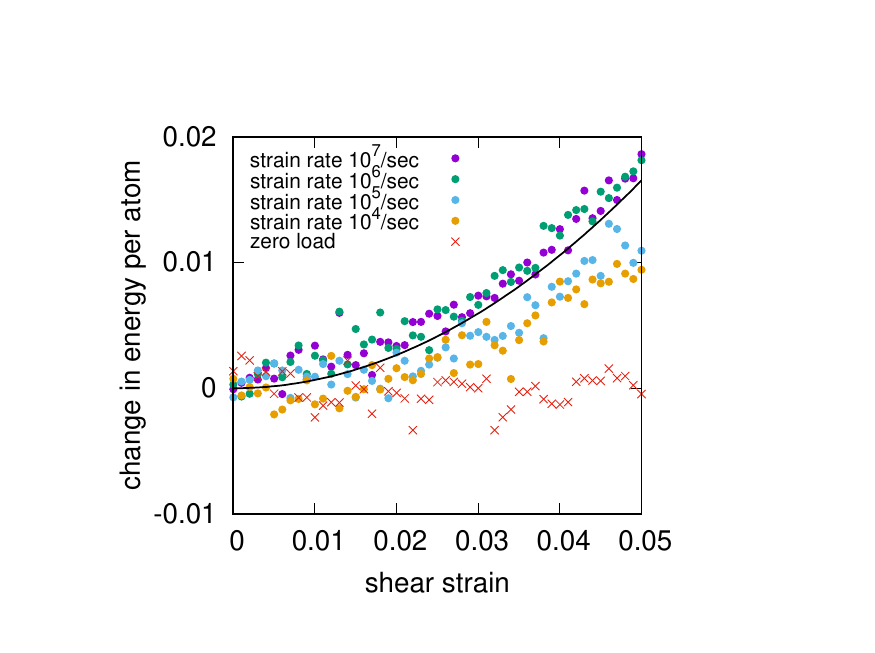}}
\subfloat[]{\includegraphics[width=0.55\linewidth,trim=1cm 0.5cm 1cm 1.5cm]{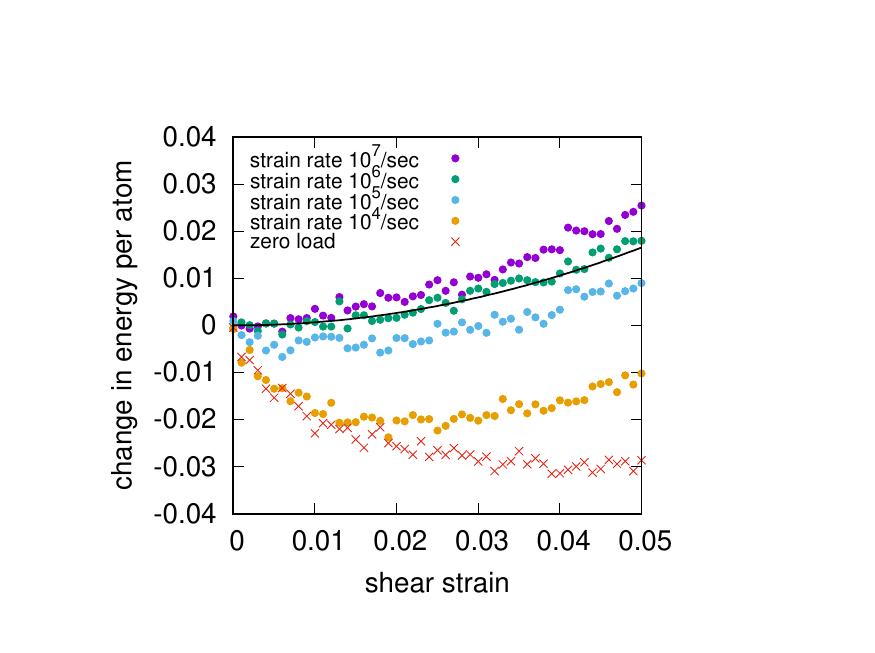}}	
\end{center}
\caption{Plot of a) shear stress and b) internal energy per atom as a function of shear strain for four orders magnitude strain rate. The shear stress is divided by the shear modulus, $G$, measured at the highest strain-rate. In b) the theoretical quadratic elastic response ($1/2G\gamma^{2}$) is also shown. Also shown is zero-load data, plotted with respect to the product of physical simulation time and the chosen strain of $10^{4}$/sec. c) and d) show the corresponding average potential energy per atom as a function of shear strain for icosahedral and non-icosahedral coordinated atoms respectively. The same color scheme is used in all panels.}
\label{figLoading}
\end{figure}

Fig.~\ref{figLoading}a shows the shear stress-strain behaviour for strain-rates spanning four orders of magnitude. The highest strain-rate, $10^{7}$/sec, is the typical nano-second deformation timescale considered in contemporary literature, whereas the lowest rate, $10^{4}$/sec, involves one micro-second per 0.01 strain increment. The figure demonstrates an initial linear response for all strain rates, and at shear strains above 0.02-0.03, an increasing strain-rate effect is seen, which leads to a strain-rate dependent peak-stress response and a transition to plastic flow. This plastic regime will be considered in a forthcoming publication. A closer inspection of the elastic region reveals a weak shear softening (reduction of effective elastic shear modulus) with decreasing strain rate. Fig.~\ref{figLoading}b plots the internal energy per atom as a function of shear-strain. Unlike Fig.~\ref{figLoading}a, a strong strain-rate dependence is now seen in the nominally elastic regime, where for the lowest strain-rate, the internal energy significantly decreases. For increasing strain-rates, the internal energy curves tend to approach the expected elastic response, $1/2G\gamma^{2}$ where $G$ is the shear modulus derived from the $10^{7}$ strain data (derived from strain values below 0.01).

Atoms are now classified into two groups as a function of the shear strain: those icosahedrally and non-icosahedrally coordinated using Voro++~\cite{Rycroft2009} within the OVITO software package~\cite{Stukowski2010}. For our well-relaxed sample, approximately 25\% of the smaller atoms are icosahedrally coordinated. A connectivity analysis of the icosahedrally coordinated atoms reveals a dominant system spanning cluster consisting of connected fragments of C15 Laves backbone polyhedra. This is a known structural feature of the model LJ glass~\cite{Pedersen2010,Derlet2020} and also of small clusters in a model CuZr glass~\cite{Zemp2014,Zemp2016,Ryltsev2016}. For each group of atoms, an average potential energy per atom is calculated as a function of shear strain. These potential energies are obtained by performing a congjugate gradient relaxation to a local inherent state configuration to remove the thermal component (referred to as cg-quenched), and thus should be distinguished by the internal energies of Figs.~\ref{figLoading}a-b which contain a thermal component. These icosahedral and non-icosahedral energies are shown respectively in Figs.~\ref{figLoading}c and d. Also shown in these figures is the expected elastic response, $1/2G\gamma^{2}$. Fig.~\ref{figLoading}c shows that the average energy of the icosahedral atoms follow more closely the elastic response at all shear-rates with only a weak dependence on strain rate, whereas Fig.~\ref{figLoading}d confirms that the non-icosahedrally coordinated atoms underlie the strain-rate dependent energy relaxation seen in Fig.~\ref{figLoading}b. The red data set in panels b, c, and d describe the evolution of a later discussed zero-load configuration (see Sec.~\ref{ssecZeroLoad}).

\begin{figure}
	\begin{center}
		\subfloat[]{\includegraphics[width=0.55\linewidth,trim=1cm 0.5cm 1.5cm 1cm]{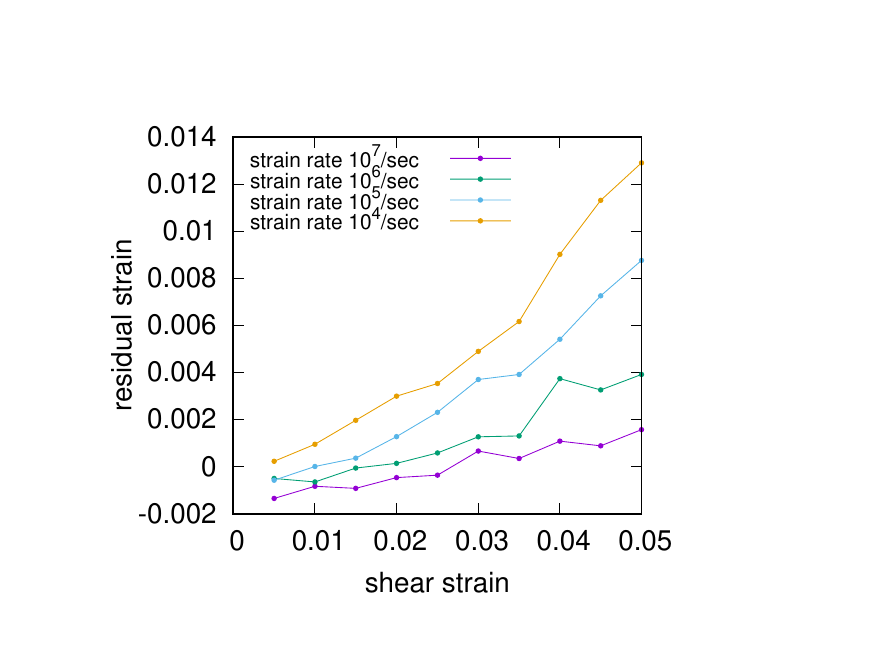}}
		\subfloat[]{\includegraphics[width=0.55\linewidth,trim=1cm 0.5cm 1.5cm 1cm]{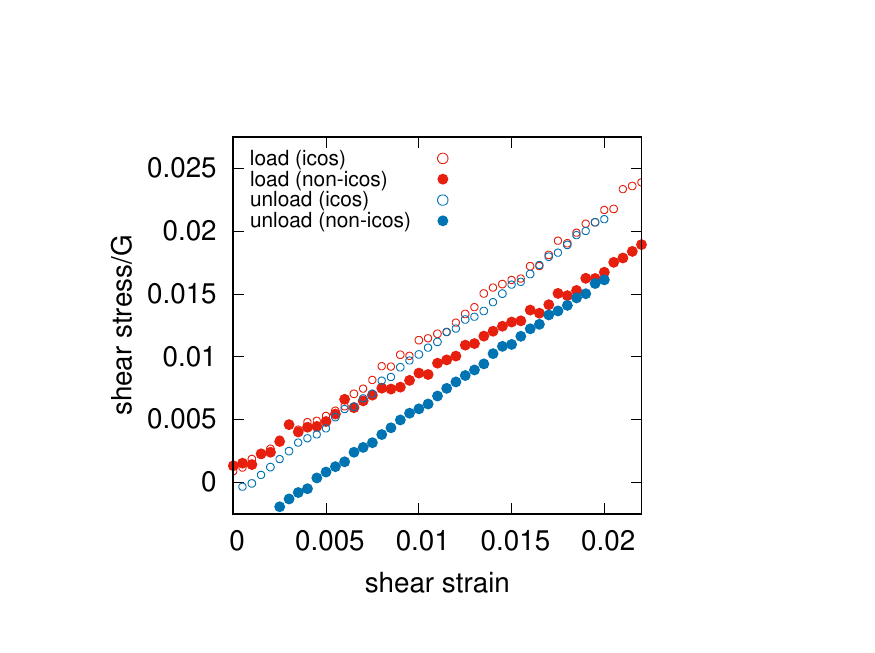}}
	\end{center}
	\caption{a) Residual strain upon unloading at different strains for the strain rates considered. b) Average shear stress response of icosahedral and non-icosahedral regions during loading and unloading of the $10^{4}$/sec strain rate deformation.}
	\label{figUnloading}
\end{figure}

To investigate the degree of micro-plasticity occurring during the deformation, samples are unloaded to zero shear stress at various strains. The unloadings were performed at the upper strain rate of $10^{7}$/sec, to ensure minimal structural relaxation and therefore maximal probing of the internal stress state at a particular strain. Fig.~\ref{figUnloading}a displays the resulting residual strain for all the strain rates. The figure reveals that micro-plasticity occurs within the nominally elastic regime for all strain rates, and increases (in terms of its residual strain signature) as the strain rate reduces. Within this elastic regime, the residual strain increases with strain and for the lowest strain rate the residual strain begins to increase more rapidly beyond a strain of 0.02 suggesting the onset of the flow regime at these micro-second timescales. The observation of micro-plasticity is compatible with the findings entailed in Fig.~\ref{figLoading}a, suggesting that the icosahedral regions respond purely elastically, whereas the non-icosahedral regions undergo some form of plasticity that surprisingly results in structural relaxation. To gain further insight into the origin of this micro-plasticity, the spatial load-bearing properties of the glassy structure are now investigated.

To determine the spatial origin of the evolving global shear stress, the virial stress $\sigma^{\mu\nu}=1/(2V)\sum_{ij}F^{\mu}_{ij}r^{\nu}_{ij}$ (where $\mathbf{F}_{ij}$ and $\mathbf{r}_{ij}$ are the force and distance vectors between atoms $i$ and $j$, $V$ is the volume of the simulation cell, and the factor of 1/2 is used to correct for double counting) is written as two terms, one involving a summation of $i$ over icosahedrally coordinated atoms and the other over non-icosahedrally coordinated atoms. These two terms are used individually and divided respectively by the fraction of icosahedrally ($\Omega_{\mathrm{I}}$) and non-icosahedrally ($\Omega_{\mathrm{NI}}$) coordinated atoms. This allows determining the average local stress for each of the two atomic environments individually. Doing so, modifies the $1/V$ pre-factor to give a representative volume of the icosahedral and non-icosahedral regions, and therefore an estimate of the corresponding stresses. An average of the two quantities, weighted according to  $\Omega_{\mathrm{I}}$ and $\Omega_{\mathrm{NI}}$, gives the exact global stress of the system. Fig.~\ref{figUnloading}b shows their evolution as a function of strain for the $10^{4}$/sec strain-rate sample derived from the corresponding cg-quenched inherent-state configurations. The data reveals that different shear stresses are carried by both regions of the material, with the non-icosahedral regions carrying a lower stress. The observed difference between the icosahedral and non-icosahedral stress responses decreases with increasing strain rate (data for other strain rates not shown), as suggested in Figs.~\ref{figLoading}c and d which show similar elastic responses at the higher strain rates. Thus the difference in shear stress between the two regions increases with decreasing strain rate, and is therefore the origin of the elastic softening seen in Fig.~\ref{figLoading}a.

\subsection{Negative creep and strain recovery} \label{ssecNegCreep}

The picture which therefore emerges is that, upon loading, the micro-plasticity occurring within the non-icosahedrally coordinated regions facilitates a reduction in stress (and energy) within these regions. Thus, upon unloading, a back-stress develops and produces the residual strains seen in Fig.~\ref{figUnloading}a. The unloaded sample therefore experiences a heterogeneous internal shear stress distribution in which the icosaheral regions are under a positive shear stress and the non-icosahedral regions are under a negative shear stress, which together result in a net-zero shear stress. The observation that an increasing residual strain is observed with decreasing strain rate is strong evidence for thermally-activated micro-plasticity. If this is indeed the case, the unloaded heterogeneous micro-structure should experience negative creep, with the non-icosahedral regions facilitating micro-plasticity to reduce the negative stress which in turn will reduce the positive stress of the non-icosahedral regions, homogenizing the internal stress to eventually exhaust the negative creep. This has the experimental implication that time-dependent strain recovery may be present after a constant force experiment within the nominally elastic regime, evidence of which can be found in Refs.~\cite{Castellero2008,Lei2019}.

\begin{figure}
	\begin{center}
		\subfloat[]{\includegraphics[width=0.55\linewidth,trim=1cm 0.5cm 1.5cm 1cm]{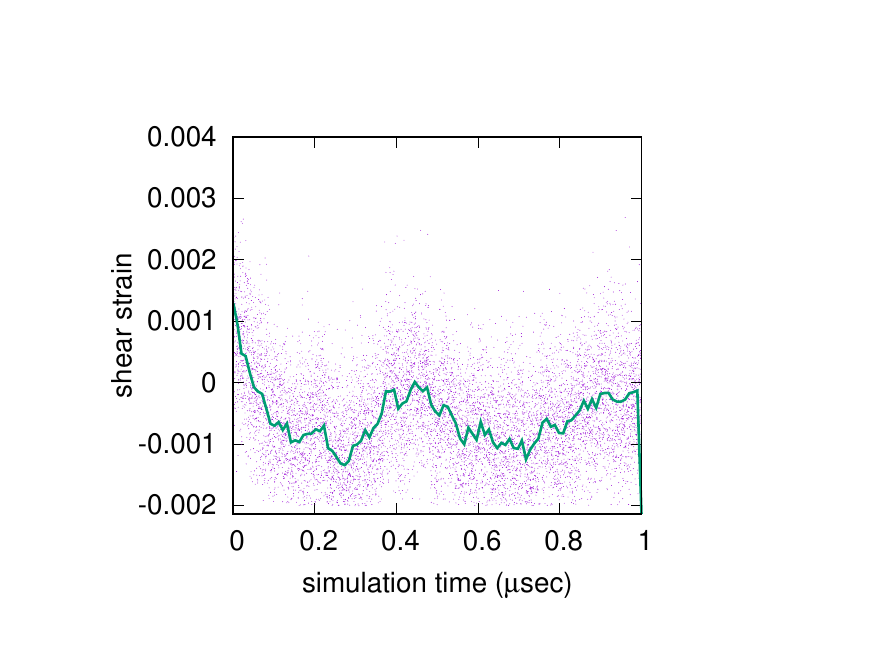}}
		\subfloat[]{\includegraphics[width=0.55\linewidth,trim=1cm 0.5cm 1.5cm 1cm]{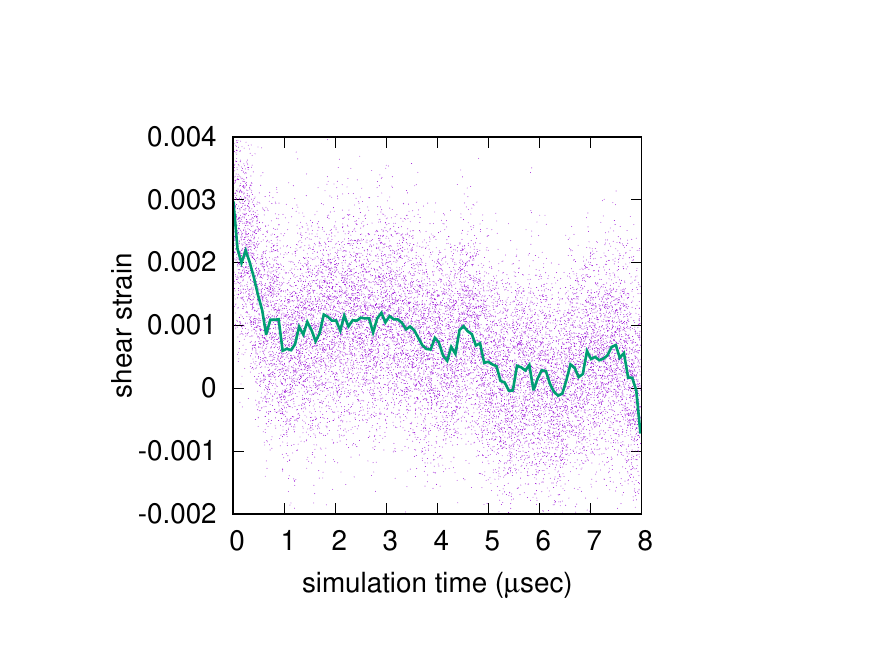}}
	\end{center}
	\caption{Strain recovery evolution of samples unloaded at a shear strain of 0.02 for the $10^{4}$/sec and $10^{5}$/sec strain rates. The green solid curves indicate smoothed data.}
	\label{figCreep}
\end{figure}

Fig.~\ref{figCreep} displays the resulting shear strain evolution as a function time of samples unloaded at a shear strain of 0.02 obtained at the two slowest strain rates a) 10$^{5}$ and b) 10$^{4}$. The NPT simulations are performed at 0.8$T^{\mathrm{NPT}}_{\mathrm{f}}$ using an additional zero shear stress barostat in the x-y plane. At $t=0$ the data begins with an initial residual strain of approximately 0.001 and 0.003 for the respective strain rates, 10$^{5}$ and 10$^{4}$, as seen in Fig.~\ref{figUnloading}a. For the 10$^{5}$ loading to 0.02 the residual strain rapidly drops and by 100 nano-seconds begins fluctuating around a negative value of approximately -0.0005 at a time scale of order 200 nano-seconds. For the 10$^{4}$ loading to 0.02 the residual strain also drops to a fluctuating value, now after several micro-seconds. In this case the fluctuations occur at the micro-second timescale. Thus in both cases, negative creep is indeed observed leading to strain recovery at a timescale larger but of similar order to of the loading to 0.02 shear strain.

\subsection{Comparison to zero-load evolution} \label{ssecZeroLoad}

The observations that 1) significant structural relaxation occurs during deformation at low enough strain rates, 2) this occurs in non-icosahedral regions which also undergo a reduction in shear, 3) residual strain exists upon unloading indicating micro-plasticity and 4) negative creep is seen resulting in recovery towards the pre-loading geometry, all point towards a strong thermally-activated micro-plasticity component.

Because of the inherently non-equilibrium nature of the glass structure, structural relaxation under zero-load also occurs via the thermal activation of microscopic scale processes. These localized structural changes are mediated by localized structural excitations (LSE). LSEs, can be observed in atomistic simulations at finite temperature within the dynamical heterogeneities of the under-cooled liquid~\cite{Donati1998,Schroeder2000,Gebremichael2004,Vogel2004,Chandler2010,Kawasaki2013}, and also well below the glass transition temperature of the amorphous solid both~\cite{Swayamjyoti2014,Derlet2017,Derlet2017a,Derlet2020}. They involve a core region in which atoms rearrange, often in a string-like fashion in which one atom replaces the position of a neighbouring atom. This localized activity is accommodated by a far-field Eschelby strain field which has a similar energy-scale as the LSE core~\cite{Swayamjyoti2016}. Thermally-activated LSEs should be distinguished from athermally stress-driven shear transformation zones~\cite{Falk1998,Maloney2004,Shi2005,Demkowicz2005} arising from inflections in the PEL giving core structures significantly more localized than that of LSEs.

\begin{figure}
	\begin{center}
		\subfloat[]{\includegraphics[width=0.55\linewidth,trim=1cm 0.5cm 1.5cm 1cm]{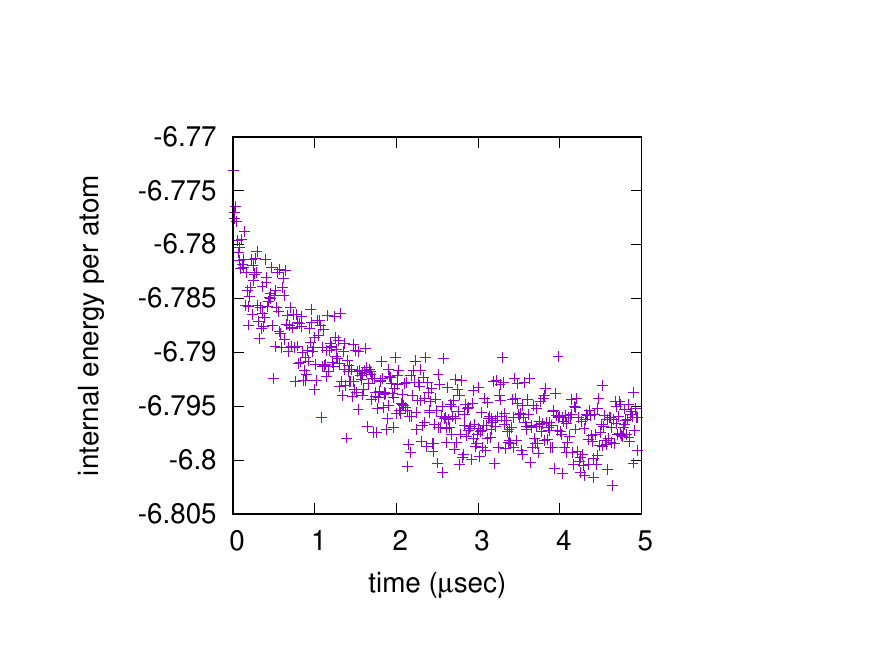}}
		\subfloat[]{\includegraphics[width=0.55\linewidth,trim=1cm 0.5cm 1.5cm 1cm]{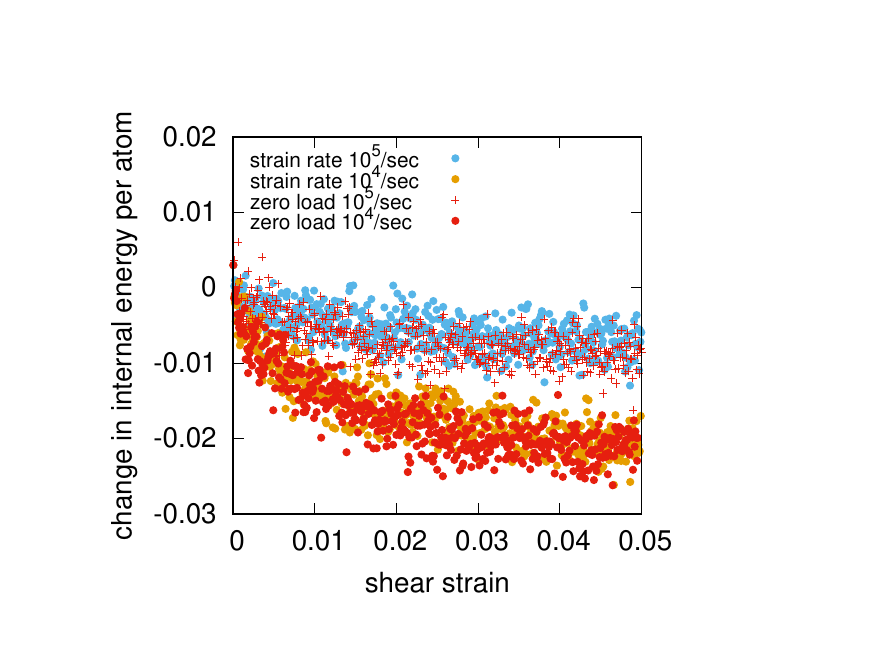}}
	\end{center}
	\caption{a) Plot of internal energy evolution of zero-load sample as a function of time. b) Plot of internal energy evolution during loading minus the elastic energy component compared to the zero load internal energy evolution shown in a), plotted as a function of $\dot{\varepsilon}\times t$.}
	\label{figZeroLoad}
\end{figure}

These considerations motivate the questions also eluded to in the introduction: i) are there differences between the structural relaxation occurring under load and the relaxation processes that would normally occur under zero load? and ii) would such zero-load structural relaxation be related to the thermally-activated micro-plasticity observed in Fig.~\ref{figLoading}?

To investigate the nature of structural relaxation without load, isothermal simulations are performed for 5 micro-seconds at zero fixed pressure and at $0.8T^{\mathrm{NPT}}_{\mathrm{f}}$, starting with the initial configuration used for the loading simulations. The cohesive energy per atom as a function of time is shown in Fig.~\ref{figZeroLoad}a showing significant structural relaxation has occurred over the time scale of 5 micro-section. Such data may also be plotted as a function of $\dot{\varepsilon}\times t$, giving a more direct indication of the structural relaxation occurring under zero load at the timescale associated with a particular strain achieved during deformation at a given strain-rate. Fig.~\ref{figZeroLoad}b plots the zero-load relaxation data with respect to $\dot{\varepsilon}\times t$ for the $10^{4}$/sec and $10^{5}$/sec strain rates, and the corresponding data of the deformation simulations with the elastic energy contribution ($1/2G\gamma^{2}$) removed. The curves show that the structural relaxation is similar. The zero load data is also shown in Figs.~\ref{figLoading}b-d, which demonstrates that when the zero-load data is partitioned into icosahedral and non-icosahedral contributions --- as under load, the non-icosahedral regions exhibit the bulk of the relaxation with the icosahedral regions experiencing minimal relaxation (see Figs.~\ref{figLoading}c-d). These simulations demonstrate the important result that in terms of energy decrease, very similar relaxation occurs for both zero load and during (and indeed somewhat beyond) the initial nominally elastic loading. 

\begin{figure}
\begin{center}
\includegraphics[width=0.55\linewidth,trim=1cm 0.5cm 1cm 1cm]{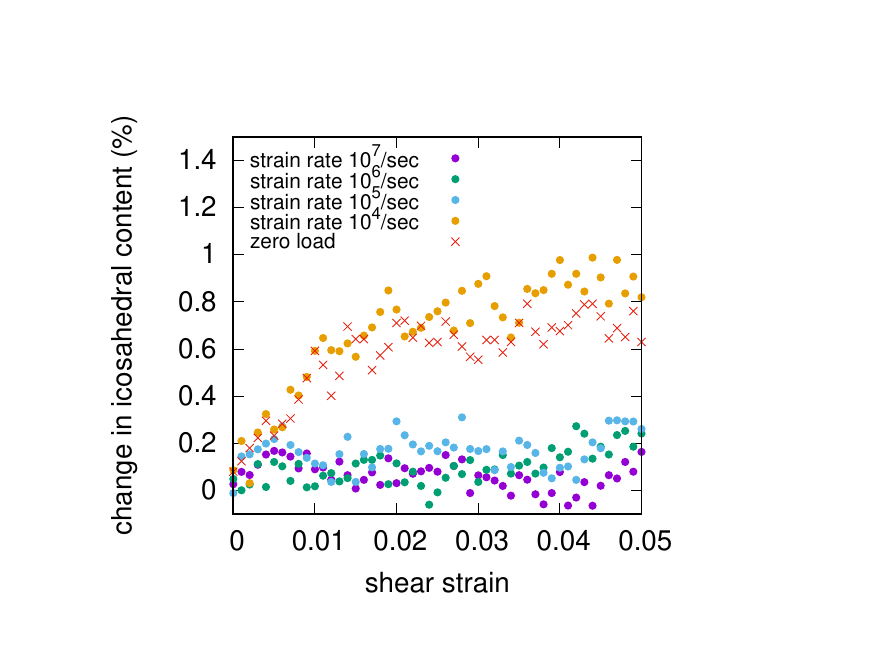}
\end{center}
\caption{a) Icosahedral content as a function of strain, spanning 4 orders of magnitude strain rate. Zero-load data is also shown as a function of the product of physical simulation time and the chosen strain of $10^{4}$/sec.}
\label{figIcos}
\end{figure}

This structural relaxation is quantified in Fig.~\ref{figIcos}, which plots the evolution of the icosahedral content during loading as a function of shear strain for all strain rates. The data has been corrected for a general decrease in icosahedral content as a function of shear strain. This smooth reduction in icosahedral content, which is reversible upon unloading, also exists in purely athermal deformation and originates from the Voronoi based detection algorithm not being able to identify some strongly strained local icoasahedral environments. For the highest $10^{7}$/sec strain rate,  Fig.~\ref{figIcos} shows that the corrected icosahedral content slightly decreases, indicating weak rejuvenation and a likely reason for why the energy per atom increases more rapidly than that expected from elasticity (see Figs.~\ref{figLoading}b and d). For the two intermediate strain rates, the icosahedral content remains approximately constant. However for the $10^{4}$/sec strain rate, the icosahedral content increases until a strain of 0.01-0.02, and then plateaus at higher strains. Also shown is the icosahedral evolution for the zero-load sample plotted as a function of $\dot{\varepsilon}\times t$ for the $10^{4}$/sec strain rate, displaying a similar evolution in icosahedral content. Thus the initial rise in icosahedral content is due to the initial transient relaxation regime seen in both zero-load and loading simulations (Fig.~\ref{figZeroLoad}). 

Returning to questions i) and ii), structural evolution in terms of energy relaxation is similar for a zero of finite applied shear stress, with both being mediated by thermally activated LSE activity. For the case of an applied load, the LSE activity clearly results in a net shear stress reduction with the non-icosahedral regions. How this similarity and difference might arise will be discussed in Sec.~\ref{secDis}.

\section{Microcopic structural analysis}
	
\subsection{Local stress measure distribution} \label{ssecLStress}

The results of the loading and unloading simulations reveal a developing internal stress heterogeneity that correlates with the icosahedral and non-icosahedral atomic environments. How is this stress heterogeneity manifested at the local microscopic level? Insight into this question may be seen by rewriting the expression for the global stress as
\begin{equation} \sigma^{\mu\nu}=\frac{1}{2V}\sum_{ij}F^{\mu}_{ij}r^{\nu}_{ij}=\frac{1}{2NV_{\mathrm{atom}}}\sum_{ij}F^{\mu}_{ij}r^{\nu}_{ij}=\frac{1}{N}\sum_{i}\sigma^{\mu\nu}_{i}, \label{EqnLocalStress}
\end{equation}
where $N$ is the number of atoms and $V_{\mathrm{atom}}$ is the average volume per atom. $\sigma^{\mu\nu}_{i}$  should be viewed as a measure of atom $i$'s contribution to the global stress, where the average value gives the global stress. It may also bee seen as an estimate of the actual local atomic stress. 

\begin{figure}
\begin{center}
\subfloat[]{\includegraphics[width=0.55\linewidth,trim=1cm 0.5cm 1cm 1.5cm]{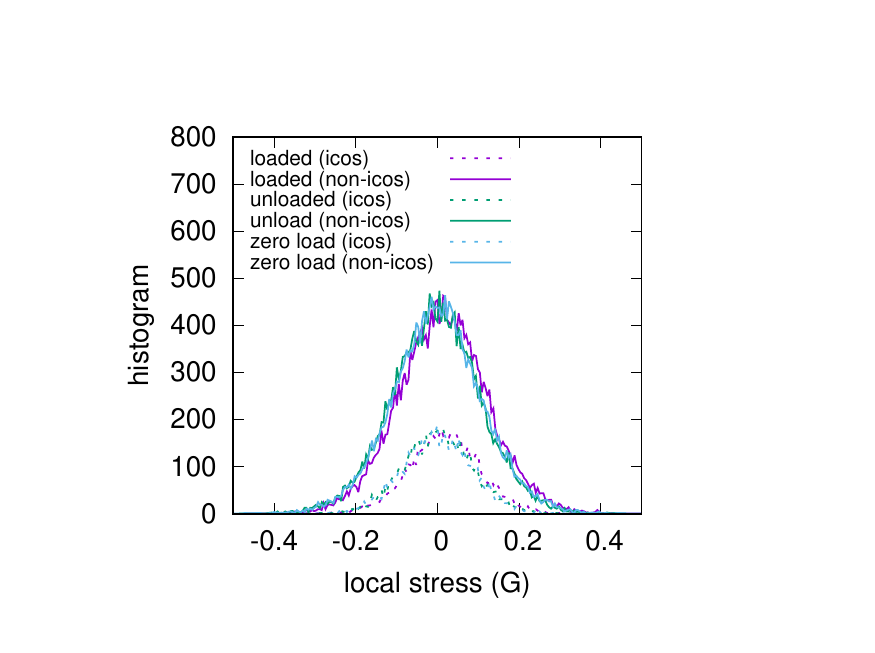}}
\subfloat[]{\includegraphics[width=0.55\linewidth,trim=1cm 0.5cm 1cm 1.5cm]{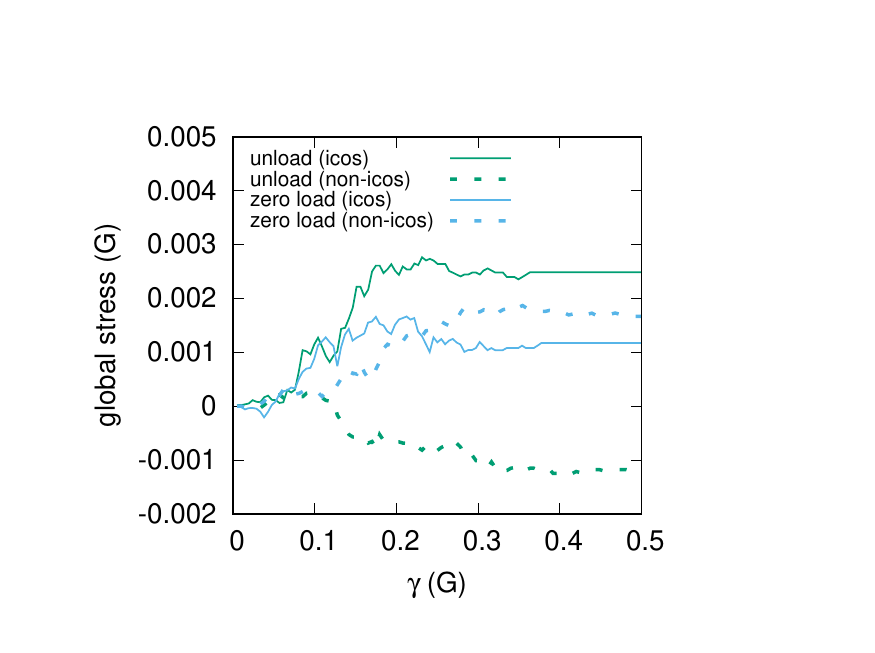}}
\end{center}
\caption{a) Histograms of the local shear-stress measure for the sample loaded to 0.02 shear strain at a strain rate of $10^{4}$/sec, the unloaded and zero load (initial prior to loading) sample. The distributions are shown for both the icoshadral and non-icosahedral atoms. b) The total global shear stress as a function of a local stress integration limit $\pm\gamma$ for the zero load and unloaded samples (see Sec.~\ref{ssecLStress}).}  
\label{figLocalStress}
\end{figure}

Fig.~\ref{figLocalStress}a displays a histogram of $\sigma^{xy}_{i}$ for the zero-load (initial) sample, the sample loaded to 0.02 shear strain at the strain rate of $10^{4}$/sec, and the subsequently unloaded sample to approximately zero $xy$ shear stress. The distributions are shown for the icosahedral and non-icosahedral atom classes. For the zero-load and unloaded samples, the distributions are centered at approximately zero, whereas for the loaded sample the distribution is shifted by approximately $0.02G$. Inspection of the figure reveals a bandwidth of shear stresses that is significantly larger than the applied stresses considered in the present work, and of a general form which is the same for both classes of atoms.  Indeed, for all three samples, the most probable regions of the distributions are well described by a shifted Gaussian, for which the width is $0.159G$ for the non-icosahedral data and $0.128G$ for the icosahedral data. For the zero-load data the non-icosahedral/icosahedra Gaussian centers are at magnitudes at and below $\sim0.0001G$, whereas for the unloaded data, the Gaussians are centered at stresses of $-0.003G$/$0.00085G$. For the loaded sample, the centers are at $0.014G$ and $0.020G$.

Fig.~\ref{figLocalStress}b shows how the global averages of the icosahedral and non-icosahedral regions are obtained in terms of the distributions shown in Fig.~\ref{figLocalStress}a. This is done for the zero-load and unloaded sample by performing the summation in Eqn.~\ref{EqnLocalStress} but only including terms within the range $\pm\gamma$, and plotting the summation as a function of $\gamma$. Fig.~\ref{figLocalStress}b demonstrates that for small values of $\gamma$, which includes only the most probable part of the distribution, the global stress remains quite small. Deviations away from this begin in the tails of the distribution (below values of approximately the full-half-width-maximum), leading to the clear differences in the stress states of the icosahedral and non-icosahedral regions of the unloaded sample. On the other hand, the zero-load sample shows comparable values for the two classes of atoms, albeit at a clearly non-zero value of the global shear stress. This again reflects the fluctuating stress state of our finite sample, discussed in Sec.~\ref{sssample}.

The results of this section indicate that the stress heterogeneity associated with the icosahedral and non-icosahedral regions manifests itself in differences in the local atomic stress distributions, where the distribution is consistently narrower for the icosahedral regions. On the other hand, the average stress difference of these regions originates from large, but rare, local stress contributions at the tails of these distributions. This result suggests that much of the internal stress state of the material remains inactive, with only certain regions undergoing structural change associated with relaxation and/or micro-plasticity.

\subsection{Localized structural excitations} \label{ssecMob}

\begin{figure}
	\begin{center}
		\includegraphics[width=0.95\linewidth,trim=3.5cm 4cm 3.5cm 2cm,clip]{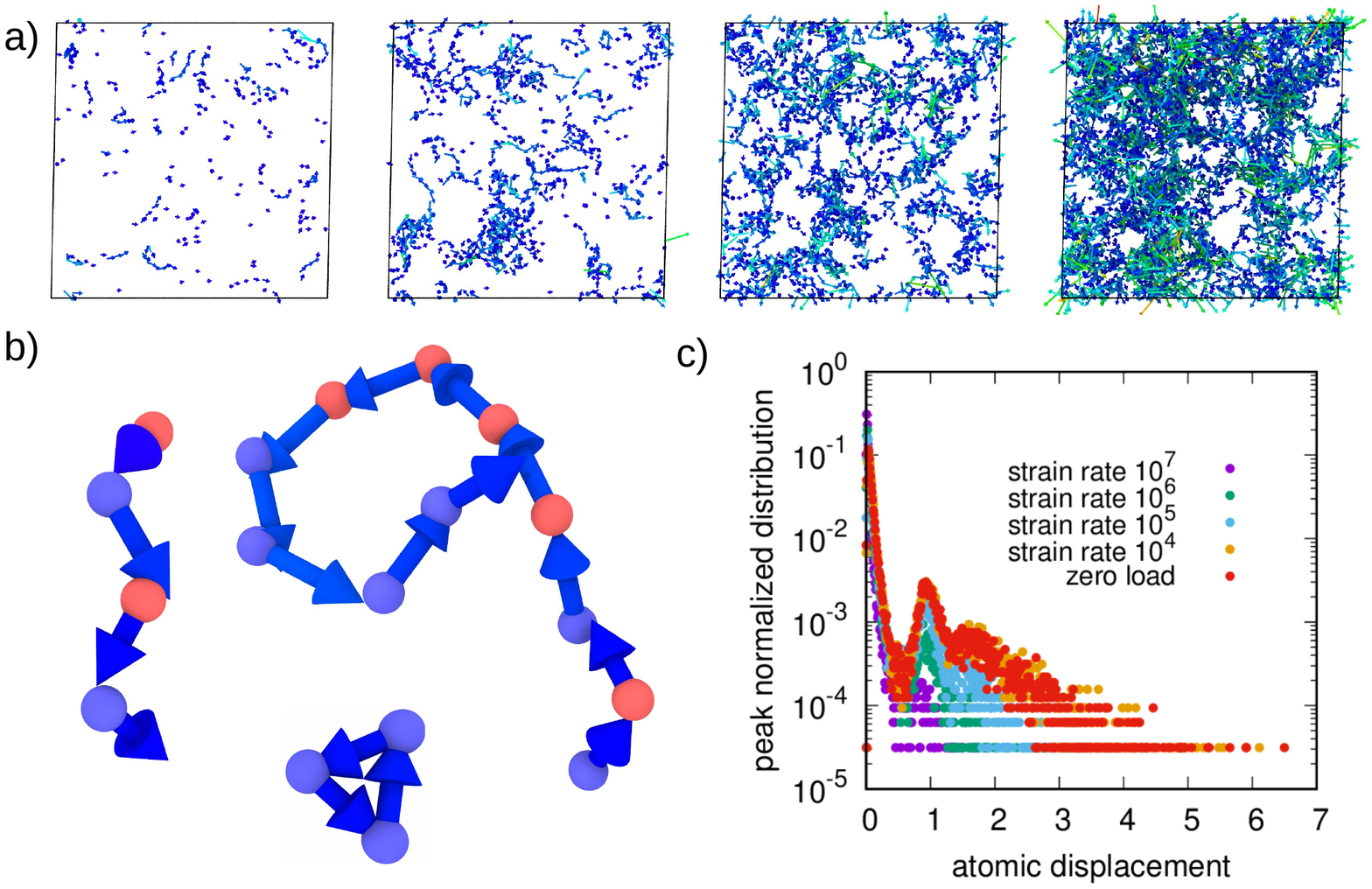}
	\end{center}
	\caption{a) Atomic displacement maps showing the LSE mediated structural activity between the configuration prior to loading and at 0.02 strain for the strains rates $10^{7}$/sec, $10^{6}$/sec, $10^{5}$/sec, and $10^{4}$/sec corresponding to the time intervals  $\Delta t$ of 2.0 nsec, 20.0 nsec, 200.0 nsec and 2 $\mu$sec. b) Examples of individual LSE activity consisting of string-like sequences of atomic displacements. c) Van hove spatial correlation function $G(r,\Delta t)$ associated with the configurations used in a). Data associated with the zero-load sample is also shown for $\Delta t=2$ $\mu$sec. }
	\label{figLES}
\end{figure}

The present data, indicates that a similar thermally-activated relaxation scenario occurs for both zero- and finite-load conditions. Fig.~\ref{figLES}a plots the atomic displacement between the sample prior to loading and the sample at 0.02 strain for the four considered strain rates corresponding to respective time intervals  of 2.0 nsec, 20.0 nsec, 200.0 nsec and 2 $\mu$sec. The displacement vectors are corrected for affine disortion due to the global shear strain. Only displacements above 0.6$\sigma$ are visualized. The panels clearly show increased structural activity during the loading to 0.02 as a function of decreasing strain rate. The displacements may be characterized as consisting of an increasing number of localised structural rearrangements with respect to decreasing strain rate. Indeed, inspection of individual re-arrangements reveal them to be the LSEs seen under zero load in past work~\cite{Derlet2017,Derlet2017a,Swayamjyoti2014,Derlet2020}. Fig.~\ref{figLES}b displays a number of such LSEs, showing the string-like extended atomic displacement sequence.

To quantitatively confirm this similarity in LSE structure, atomic displacements occurring between a time interval, $\Delta t$, are binned to obtain the van Hove self-correlation function, $G(r,\Delta t)=\langle\delta(r-|\mathbf{r}_{i}(\Delta t)-\mathbf{r}_{i}(0)|)\rangle$, a quantity that is experimentally accessible using coherent photon correlation spectroscopy methods~\cite{Das2019}. Fig.~\ref{figLES}c plots the resulting  correlation function obtained from the configurations prior to load and at a shear strain of 0.02 for all strain rates corresponding to a respective $\Delta t$ of 2.0 nsec, 20.0 nsec, 200.0 nsec and 2 $\mu$sec. All correlation functions demonstrate two structural features: 1) a narrow peak whose maximum is close to zero with a width that spans displacements below the characteristic bond-length (here $\sim1\sigma$), and 2) a peak at this characteristic bond-length which exhibits an exponential tail at larger displacements. The intensity of this latter peak increases with decreasing strain rate (increasing $\Delta t$). Fig.~\ref{figLES}c also includes the van Hove self-correlation function at zero-load for $\Delta t=2\mu$sec which due to a similar time interval may be directly compared to the $10^{4}$ strain rate data, and which demonstrates a close overlap with the lowest strain rate. A similar overlap can be seen for the higher strain rates when using the appropriate $\Delta t$ (not shown). 

In addition, comparison of the second peak associated with bond-length displacements, for the different strain rates show a similar overall structure --- the main effect being that the intensity of the second peak increases with decreasing strain rate. A similar effect is seen for the correlation functions derived from the icosahedral and non-icosahedral regions of the sample. Here the atomic diplacement statistics is again similar in form, but the intensity of the second peak is larger form the non-icosahedral regions. Thus, between the two spatial regions, only the level of activity is different not the nature of the activity, where for the $\Delta t=2\mu$sec correlation functions, only $\approx4\%$ of icosahedrally coordinated atoms were involved in bond-length scale displacements, whereas for non-icosahedrally coordinated atoms it was $\approx16\%$.

More generally, the character of thermally-activated LSE activity is the same, whether under load or not, where the second peak is associated with LSE core atom bond length (or greater) displacements. The observed exponential tail at larger distances is a characteristic of collective behaviour and conjectured to be a universal feature of the glass transition regime~\cite{Chaudhuri2007}. The present work extends this property to the thermally-activated LSE processes below the glass transition regime. The sub-bond-length displacements associated with the first peak correspond to the accommodating LSE Eschelby strain field indicating an intrinsic landscape of small barrier energies which has been viewed as a collective generalization of rattling-in-the-cage excitations~\cite{Ciamarra2016}.

\subsection{Spatial visualization}

\begin{figure}
	\begin{center}
		\includegraphics[width=0.95\linewidth,trim=2cm 6.5cm 2cm 6cm,clip]{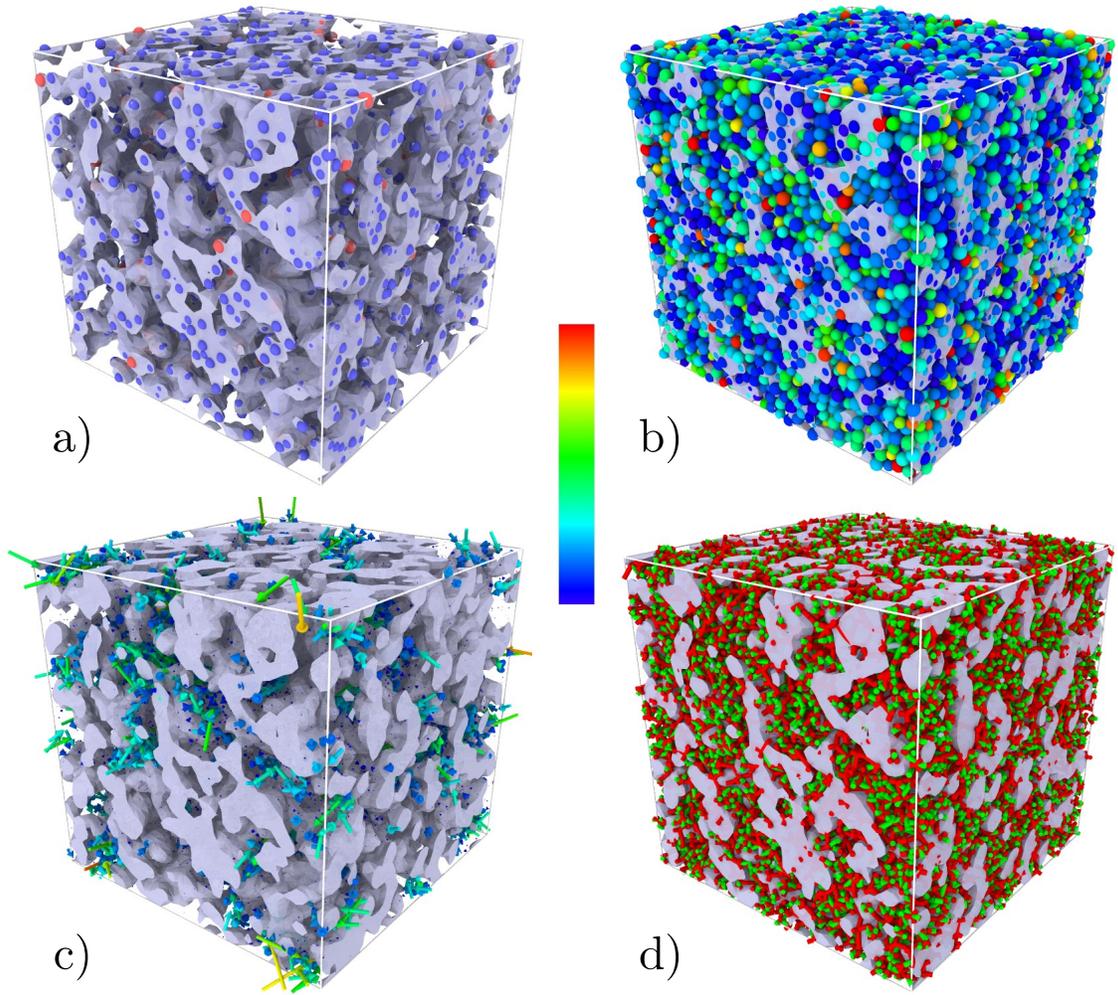}
	\end{center}
	\caption{Visualization of initial structure in which a) only atoms that are icosahedrally coordinated are shown (blue/red represent the small/large atoms) , b) all atoms are shown and coloured according to their change in Von Mises stress at 0.02 shear strain (where blue represents no change and red represents a change of greater than 0.4$G$), d) the displacement vectors between the initial configuration and the configuration at 0.02 shear strain are shown. Each vector is coloured according to its magnitude (where blue represents zero magnitude and red a magnitude of 6$\sigma$), d) the initial disclination network in which green repesents 4-fold bonds and red 6-fold bonds (See Ref.~\cite{Derlet2020a}). In all figures the visualized surface mesh encloses the atoms of a). For b) and c) intermediate stresses and displacement magnitudes can be determined by the central colour bar.}
	\label{figVis}
\end{figure}

To visualize the spatial region defined by the icosahedrally coordinated atoms, the initial configuration is shown in Fig.~\ref{figVis}a, which displays only the icosahedral atoms. Here, blue/red atoms correspond to the small/large atoms. A surface mesh encloses these atoms. Inspection of the figure demonstrates the dominance of the smaller atom and that the connected cluster of atoms is system spanning. Fig.~\ref{figVis}b plots all atoms, now coloured according to the change in local Von Mises stress arising for 0.02 strain, with the yellow-orange-red atoms indicating large changes in stress of the order of 0.4$G$. These mainly occur in the non-icosahedral regions and are relatively rare amongst largely inactive regions. This reflects the strongly heterogenous changes in internal stress at the atomic scale, revealed in Sec~\ref{ssecLStress}. Fig.~\ref{figVis}c now displays the atomic displacement vectors between this initial configuration and the configuration at 0.02 shear strain for the slowest strain rate 10$^{4}$. The displacement vectors are coloured according to their magnitude. Detailed inspection of this figure reveals bond-length displacements mainly happen within the non-icosahedral regions. These displacements occur via multiple LSE activity as discussed in Sec.~\ref{ssecMob} (see also Fig.~\ref{figLES}). Some non-icosahedral regions are also displacement-inactive reflecting the strongly heterogeneous nature of LSE activity.

To gain further insight into the structural nature of the non-icosahedral regions, a local SU(2) topological analysis is performed, following a similar approach as that done in Ref.~\cite{Derlet2020a}. This procedure, uses a modified radical Voronoi tessellation procedure to determine the bond-order of each nearest neighbour pair, defined as the number of common neighbours of the two atoms. The relevance of this description of glassy structure has its origins in Refs~\cite{Frank1952,Chaudhari1978,Nelson1983a,Nelson1983b}, where the liquid like 5-fold bond order represents the minimally frustrated packing of atoms around the bond. As a result, this bond is referred to as a defect free bond. On the other hand, 4-fold and 6-fold bonds are not minimally frustrated atomic packings, and are referred to as defected bond orders. From this perspective, the icosahedral environment is an atom with 12 defect-free bonds and thus represents a minimally frustrated short-range-order structure. The geometrical relationship between such defected bonds, how they may occur and be arranged in terms of bond orientation, has been predicted in the work of Nelson~\cite{Nelson1983a,Nelson1983b} which demonstrated via an SU(2) algebraic structure, that only certain combinations of such defect bonds could exist at a particular atomic site. The recent work of Ref.~\cite{Derlet2020a} shows that a large proportion of atomic environments of the model Lennard-Jones binary glass presently considered follow the rules dictated by the SU(2) algebraic structure.

Fig.~\ref{figVis}d displays the disclination network constructed from showing only the green (4-fold) and red (6-fold) bonds. The direction along the disclination line represents an axis on which the 5-fold local icosahedral symmetry is broken either by a 4-fold or 6-fold symmetry. Inspection of the figure reveals that within the icosahedral region, mainly 6-fold bonds occur indicating Frank-Kasper~\cite{Frank1958} SU(2) topologies, whereas in the non-icosahedral regions both 4-fold and 6-fold disclinations exist indicating so-called Nelson SU(2) topologies.~\cite{Derlet2020a}. These later topologies entail disclination nodes in which both green and red disclinations intersect according to the rules of the underlying SU(2) algebra (see Refs.~\cite{Nelson1983a,Nelson1983b}). Their presence indicates highly bond-frustrated environments. Indeed the work of Ref.~\cite{Derlet2020a} found that minimization of 4-fold disclination content indicates a reduction in the bond frustration of the glassy structure (See Fig.~5 of Ref.~\cite{Derlet2020a} and the related discussion) and that such environments are characterized by higher free volume and locally softy (possibly negative) elastic stiffness shear moduli, both of which facilitate LSE activity~\cite{Derlet2020a}. 
	
\section{Discussion} \label{secDis}

The present atomistic simulation work, which employs samples relaxed and deformed at the micro-second timescale, has revealed the emergence of thermally-activated micro-plasticity within the elastic regime as the strain rate reduces. This micro-plasticity is heterogeneously distributed in regions characterized by enhanced geometric frustration, whereas regions characterized by minimal geometrical frustration represented by a high local icosahedral content deform almost entirely elastically. This heterogeneous micro-plasticity induces a corresponding heterogeneity in the internal stress, which upon unloading, drives negative creep to homogenize the internal stress and entirely recover the residual strain induced by the micro-plasticity. The thermally-activated local structural excitations (LSEs) underlying these deformation mechanisms also results in the relaxation of the glassy structure. Such relaxation also occurs without load, but without the developing internal stress heterogeneity and may be understood from a classical thermal-activation picture of plasticity, as was applied by Spaepen~\cite{Spaepen1977}, Argon~\cite{Argon1979} and Bulatov and Argon~\cite{Bulatov1994a} which all rely on the stochastic occurrence of thermally-activated localised plastic events, biased by the applied load. These works, in turn, exploit the seminal work of Eshelby~\cite{Eshelby1957}. 

Most generally, a change in energy due to thermal activation of an LSE may be written as a sum of a core contribution, $E_{\mathrm{core}}$, and an Eshelby elastic energy contribution~\cite{Eshelby1957}:
\begin{equation}
1/2V_{0}\varepsilon^{\mu\nu}_{\mathrm{T}}\left(\sigma^{\mu\nu}_{\mathrm{T}}-\sigma^{\mu\nu}_{\mathrm{C}}-\sigma^{\mu\nu}_{\mathrm{L}}\right).
\end{equation}
Here $V_{0}$ is the volume of the Eshelby inclusion appropriate for the LSE, $\sigma^{\mu\nu}_{\mathrm{T}}$ and $\varepsilon^{\mu\nu}_{\mathrm{T}}$ are the rigid far-field elastic stress and strain due to the Eshelby construction, $\sigma^{\mu\nu}_{\mathrm{C}}$ is the correction due to internal relaxation around the Eshelby inclusion, and $\sigma^{\mu\nu}_{\mathrm{L}}$ is the characteristic external stress at the LSE. Due to its strongly local character, $E^{\mathrm{core}}$ is expected to have a much weaker dependence on $\sigma^{\mu\nu}_{\mathrm{L}}$ than the Eshelby contribution. Taking $\sigma^{\mu\nu}_{\mathrm{L}}$ as a simple shear in the $xy$ plane of magnitude $\sigma$, it's contraction with $\varepsilon^{\mu\nu}_{\mathrm{T}}$ gives $\varepsilon^{\mu\nu}_{\mathrm{T}}\sigma^{\mu\nu}_{\mathrm{L}}=2\sigma\gamma_{\mathrm{T}}\cos2\theta_{\mathrm{T}}$ and an energy which depends on the LSE's Eshelby shear-strain magnitude $\gamma_{\mathrm{T}}$ and orientation  $\theta_{\mathrm{T}}$ within the $xy$ plane. A similar expression for the corresponding saddle-point energy may also be obtained~\cite{Bulatov1994a} resulting in a barrier energy whose leading order stress dependence will be $\sim\sigma\gamma_{\mathrm{T}}\cos2\theta_{\mathrm{T}}$, giving the general result that thermally-activated LSEs with a far-field strain signature that reduces the applied load (i.e. $\cos2\theta_{\mathrm{T}}<0$) are more likely to occur, reducing the global elastic energy through the global plastic strain increment $\varepsilon^{\mu\nu}_{\mathrm{T}}$ (here $\gamma_{\mathrm{T}}$). This picture of a barrier energy dependent on the applied stress has been confirmed through ARTn simulations which investigated how particular barrier energies varied with load, finding that in terms of a shear load, LSE barrier energies would either increase or decrease depending on their far field shear-stress signature~\cite{Swayamjyoti2016}.

The above heuristic picture of thermally-activated plasticity gives a simple reason as to why similar degrees of energy relaxation and displacement activity are seen when comparing the loaded sample to the zero-load sample --- a certain degree of thermal activation will occur irrespective of the presence of an applied load. Under zero load, over time, this activity will result in structural relaxation with no net deformation of the sample. However with an applied load, thermal activity will be biased towards those LSEs which reduce the applied elastic load. The observation that this bias does not affect the degree of thermally-activated structural relaxation (over a given time interval), infers that the glass structure is sufficiently complex to support a very large number of LSEs with a broad range of far-field strain signatures. 

The observation of a developing heterogeneity and how this can affect plasticity has been investigated in past simulation work~\cite{Baumer2013}. Indeed, in the work of Baumer and Demcowitz~\cite{Baumer2013}, which considered a model CuNb binary glass at a concentration away from its eutectic, and therefore with strong chemical segregation, the authors found a system-spanning icosahedral network at the interface between the regions rich in Cu and rich in Nb. This occurred at the the glass transition temperature regime, giving a strong analogy to the general phenomenon of gelation. Detailed analysis of their shear loading simulations performed at the glass transition temperature and at a strain rate of $10^{9}$/sec revealed the non-icosahdral regions (referred to as liquid-like) plasticised. This resulted in a characteristic cohesive energy that did not depend strongly on the applied strain, whereas the icosahedral network responded approximately elastically. Moreover, upon unloading, no residual strain was observed indicating the plastic processes within the liquid-like non-icosahedral regions were entirely reversible. The present work establishes a related phenomenon but at lower temperatures and slower rates. In particular, the significantly enhanced time scales enables the observation of recovery via classical thermally-activated micro-plasticity and creep.

This is nicely demonstrated in Fig.~\ref{figCreep}, where time-dependent strain recovery (negative creep) proceeds after unloading. Due to the slow loading rate and the ability of the non-icosahedral network to admit thermally-activated structural micro-plasticity, an incompatibility stress (and therefore strain) between both structural components emerges upon unloading. The magnitude of the residual strain must, if true thermally-activated micro-plasticity occurs during loading, depend on temperature, strain rate, and the initial structural state of the glass. These observations provide an atomistic understanding of various experimental observations of inelasticity and strain recovery after both static and cyclic loading conditions at stresses well below yielding~\cite{Lee2008,Castellero2008,Wu2009,Ye2010,Lei2019}, which typically are described with viscoelastic Kelvin-Voigt or Maxwell models. The applicability of such phenomenological laws to the here observed atomic-scale strain recovery at the micro-second time scale will be reported elsewhere.

After the elastic energy component has been subtracted from the cohesive energy, the degree of energy relaxation of the loading and zero-load curves is quite similar for the lower strain rates (Fig.~\ref{figZeroLoad}b). From this, and the observation of strain recovery, one might conclude that little relative rejuventation occurs due to the micro-plasticity. Experimental changes in enthalpy due to rejuventation occurring within the nominally elastic regime are typically a few 100 J/mol (see Refs.~\cite{Lee2008,Greer2016,Zhang2017}) and can be as high as $\sim1$ kJ/mol~\cite{Kuechemann2017}. 100J/mol is approximately 0.001 eV per atom and assuming the Lennard Jones parameter $\varepsilon$ is at the scale of one electron volt, detecting such relative energy changes in the simulation data becomes difficult because of the fluctuations associated with the finite size of the system. It is however encouraging that the present simulations could entail relative rejuvenation energies not much larger than this typical energy scale of 100 J/mol. In terms of absolute change (not relative to the zero-load sample) significant relaxation has of course occurred over the timescale of the deformation, which is certainly not the case during a typical experimental deformation. Inspection of Fig.\ref{figZeroLoad}b reveals a structural relaxation involving a reduction in energy of about 0.02$\varepsilon$ per atom. This corresponds to $\sim2$ kJ/mol when $\varepsilon$ is taken as an electron volt, which is comparable to what is seen in the fast quenching of a ribbon~\cite{Battezzati1984}. Together, these aspects motivate future deformation simulations on more relaxed configurations. For example, the configuration at 5 $\mu$sec in Fig.~\ref{figZeroLoad}a for which all $\mu$sec transients have disappeared appears as an ideal candidate for such an endeavor.

Beyond a shear strain of 0.02, Fig.~\ref{figLoading}a indicates a greater strain rate dependence with an increased shear softening with decreasing strain rate. In addition Fig.~\ref{figUnloading}a shows that with respect to shear strain, this regime entails a more rapid increase of the residual strain upon unloading. These observations, at the micro-second timescale, could be a very early signature of the onset of plastic flow at the shear strain levels where bulk metallic glasses characteristically deform in experiments.

\section{Concluding summary}

The nominal elastic regime has been studied for a well known fragile model binary glass using strain rates spanning four orders of magnitude. The work finds:
\begin{enumerate}
	\item a robust elastic response with minimal softening for all strain rates considered, despite large differences in the degree of structural relaxation occurring during the loading.
	\item the underlying reason for this is a system spanning spatial region which responds elastically, with the remaining parts of the material experiencing structural relaxation and micro-plasticity. These latter processes are responsible for the observed minimal elastic softening.
	\item upon unloading, a residual strain remains that fully recovers through negative creep.
	\item this strain recovery is driven by a back-stress within the elastically responding spatial regions, eventually leading to the homogenization of the internal stress.
	\item zero-load simulations reveal similar structural relaxation, indicating very similar thermally activated processes underlie relaxation and micro-plasticity
	\item the microscopic structural features distinguishing the elastic region are a high density of icosahedral coordinated environments and a negligible 4-fold disclination content, both of which indicate a minimally bond frustrated system.
	\item thermally activated localized structural excitations (LSEs), mediate both the relaxation and micro-plasticity, and occur mainly in the non-icosaheral regions as evidenced by enhanced atomic displacement during loading and increased disclination content (strong bond frustration).
	\item these LSEs result in large but very local changes in internal stress, that are in fact statistically quite rare, reflecting an underlying spatio-temporal heterogeneity that emerges at the micro-second time scale.
\end{enumerate}

The above should motivate new experiments on bulk metallic glasses that specifically probe the temperature, time and strain rate dependence of micro-plasticity and its relation to rejuvenation, as well as further atomistic simulations done at micro-second (and indeed the sub-millisecond) timescales. As pointed out in a recent overview article by the present authors~\cite{Maass2018}, micro-plasticity probes the fundamental thermally activated unit-scale plastic processes that are characteristic of the zero-load structural state. A proper understanding of these local processes is essential when entering the high-stress regime of collective phenomena that ultimately lead to rejuvenation, yield and plastic flow. 

\section{Acknowledgements}

The present work was supported by the Swiss National Science Foundation under Grant No. 200021-165527.


\begin{thebibliography}{10}

\bibitem{Sun2016}
Y. Sun, A. Concustell, A.L. Greer, Thermomechanical processing of metallicglasses: extending the range of the glassy state, Nat. Rev. Mater. 1 (2016), 16039.

\bibitem{Hufnagel2016}
T.C. Hufnagel, C.A. Schuh, M.L. Falk, Deformation of metallic glasses: recent developments in theory, simulations, and experiments, Acta Mater. 109 (2016), 375 (2016).

\bibitem{Tian2012}
L. Tian, Y.-Q. Cheng, Z.-W. Shan, J. Li, C.-C. Wang, X.-D. Han, J. Sun, E. Ma. Approaching the ideal elastic limit of metallic glasses, Nat. Comm. 3 (2012) 609.	

\bibitem{Wu2008}
W. F. Wu, Y. Li, and C. A. Schuh, Strength, plasticity and brittleness of bulk metallic glasses under compression: statistical and geometric effects, Philos. Mag. 88 (2008), 71-89.		

\bibitem{Harmon2007}
J.S. Harmon, M.D. Demetriou, W.L. Johnson, K. Samwer. Anelastic to plastic transition in metallic glass-forming liquids, Phys. Rev. Lett. 99 (2007), 135502.	

\bibitem{Zhu2017}
F. Zhu, A. Hirata, P. Liu, S. Song, Y. Tian, J. Han, T. Fujita, M. Chen. Correlation between Local Structure Order and Spatial Heterogeneity in a Metallic Glass, Phys. Rev. Lett. 119 (2017) 215501.		

\bibitem{Liu2018}
C. Liu, R. Maass. Elastic Fluctuations and Structural Heterogeneities in Metallic Glasses, Advanced Functional Materials 28 (2018) 1800388.		

\bibitem{Wang2018}
N. Wang, J. Ding, F. Yan, M. Asta, R.O. Ritchie, L. Li. Spatial correlation of elastic heterogeneity tunes the deformation behavior of metallic glasses,  npj Comput. Mater. 4 (2018) 19.	

\bibitem{An2016}
Q. An, K. Samwer, M.D. Demetriou, M.C. Floyd, D.O. Duggins, W.L. Johnson, W.A. Goddard. How the toughness in metallic glasses depends on topological and chemical heterogeneity, Proc. Nat. Acad. Sci. 113  (2016), 7053-7058.		

\bibitem{Zhu2016}
F. Zhu, H.K. Nguyen, S.X. Song, D.P.B. Aji, A. Hirata, H. Wang, K. Nakajima, M.W. Chen. Intrinsic correlation between [beta]-relaxation and spatial heterogeneity in a metallic glass, Na. Comm. 7 (2016) 11516.		

\bibitem{Maass2018}
R. Maass and P. M. Derlet, Micro-plasticity and recent insights from intermittent and small-scale plasticity, Acta Mater. 143 (2018), 338-363.

\bibitem{Wu2009}
Y. Wu, H.X. Li, G.L. Chen, X.D. Hui, B.Y. Wang, Z.P. Lu, Nonlinear tensile deformation behavior of small-sized metallic glasses, Scripta Materialia 61(6) (2009) 564-567.

\bibitem{Castellero2008}
A. Castellero, B. Moser, D.I. Uhlenhaut, F.H. Dalla Torre, J.F. Loffler, Room-temperature creep and structural relaxation of Mg-Cu-Y metallic glasses, Acta Materialia 56(15) (2008) 3777-3785.

\bibitem{Ye2010}
J.C. Ye, J. Lu, C.T. Liu, Q. Wang, Y. Yang, Atomistic free-volume zones and inelastic deformation of metallic glasses, Nature Materials 9(8) (2010) 619-623.

\bibitem{Lee2008}
S.-C. Lee, C.-M. Lee, J.-W. Yang, J.-C. Lee, Microstructural evolution of an elastostatically compressed amorphous alloy and its influence on the mechanical properties, Scripta Materialia 58(7) (2008) 591-594.

\bibitem{Greer2016}
A.L. Greer, Y.H. Sun, Stored energy in metallic glasses due to strains within the elastic limit, Philosophical Magazine 96(16) (2016) 1643-1663.

\bibitem{Lei2019}
T.J. Lei, L.R. DaCosta, M. Liu, W.H. Wang, Y.H. Sun, A.L. Greer, M. Atzmon, Microscopic characterization of structural relaxation and cryogenic rejuvenation in metallic glasses, Acta Materialia 164 (2019) 165-170.

\bibitem{Ross2017}
P. Ross, S. Kuechemann, P.M. Derlet, H. Yu, W. Arnold, P. Liaw, K. Samwer, R. Maass, Linking macroscopic rejuvenation to nano-elastic fluctuations in a metallic glass, Acta Materialia 138 (2017) 111-118.

\bibitem{Zhang2018}
P. Zhang, J.J. Maldonis, Z. Liu, J. Schroers, P.M. Voyles, Spatially heterogeneous dynamics in a metallic glass forming liquid imaged by electron correlation microscopy, Nature Communications 9(1) (2018) 1129.

\bibitem{Ruta2017}
B. Ruta, E. Pineda, Z. Evenson, Relaxation processes and physical aging in metallic glasses, Journal of Physics: Condensed Matter 29(50) (2017) 503002.

\bibitem{Das2019}
A. Das, P.M. Derlet, C. Liu, E.M. Dufresne, R. Maass, Stress breaks universal aging behavior in a metallic glass, Nature Communications 10(1) (2019) 5006.

\bibitem{Das2020}
A. Das, E.M. Dufresne, R. Maass, Structural dynamics and rejuvenation during cryogenic cycling in a Zr-based metallic glass, Acta Materialia 196 (2020) 723-732.

\bibitem{Kuechemann2017}
S. Kuechemann, R. Maass, Gamma relaxation in bulk metallic glasses, Scripta Materialia 137 (2017) 5-8.

\bibitem{Frank1952}
C. Frank, Super cooling of liquids, Proc. R. Soc. A 215 (1952), 43-46.		

\bibitem{Chaudhari1978}
P. Chaudhari , D. Turnbull , Structure and properties of metallic glasses, Science 199 (1978) 11-21.		

\bibitem{Nelson1983a}
D.R. Nelson , Liquids and glasses in spaces of incommensurate curvature, Phys. Rev. Lett. 50 (1983) 982-985.			

\bibitem{Nelson1983b}
D.R. Nelson , Order, frustration, and defects in liquids and glasses, Phys. Rev. B 28	(1983) 5515-5535.			

\bibitem{Sheng2006}
H.W. Sheng W.K. Luo, F.M. Alamgir, J.M. Bai, E. Ma, Atomic packing and short-to-medium-range order in metallic glasses, Nature 439 (2006), 419-425.		

\bibitem{Ding2014}
J. Ding, Y.-Q. Cheng and E. Ma, Full icosahedra dominate local order in Cu$_{64}$Zr$_{34}$ metallic glass and supercooled liquid, Acta Mater. 69 (2014), 343-354.	

\bibitem{Ding2017}
J. Ding and E. Ma, Computational modeling sheds light on structural evolution in metallic glasses and supercooled liquids, npj Comput. Mater. 3 (2017), 9.

\bibitem{Derlet2017a}
P.M. Derlet, R. Maass. Thermally-activated stress relaxation in a model amorphous solid and the formation of a system-spanning shear event, Acta Mater. 143 (2018), 205-213.		

\bibitem{Derlet2018}
P. M. Derlet, R. Maass, Local volume as a robust structural measure and its connection to icosahedral content in a model binary amorphous system, Materialia 3 (2018), 97-106. 		

\bibitem{Derlet2020}
P. M. Derlet, R. Maass, Emergent structural length scales in a model binary glass --- the micro-second molecular dynamics time-scale regime, J. Alloys. Comp. 821 (2020) 153209

\bibitem{Derlet2020a}
P. M. Derlet, Correlated disorder in a well relaxed model binary glass through a local SU(2) bonding topology, under review (2020).arXiv:2007.08878		

\bibitem{Ma2015}
E. Ma, Tuning order into disorder, Nat. Mater. 14 (2015) 547-552.		

\bibitem{Falk1998}
M. L. Falk and J. S. Langer, Dynamics of viscoplastic deformation in amorphous solids, Phy. Rev. E 57 (1998) 7192-7205. 		

\bibitem{Maloney2004}
C. Maloney, A. Lema\^{i}tre. Subextensive scaling in the athermal, quasistatic limit of amorphous matter in plastic shear flow, Phys. Rev. Lett. 93 (2004), 016001.	

\bibitem{Demkowicz2005}
M.J. Demkowicz, A.S. Argon. Liquidlike atomic environments act as plasticity carriers in amorphous silicon, Phys. Rev. B 72 (2005) 245205.			

\bibitem{Spaepen1977}
F. Spaepen. Microscopic Mechanism for Steady-State Inhomogeneous Flow in Metallic Glasses, Acta Metall. 25 (1977) 407-415.			

\bibitem{Argon1979}
A.S. Argon. Plastic Deformation in Metallic Glasses, Acta Metall. 27 (1979) 47-58.		

\bibitem{Bulatov1994a}
V.V. Bulatov and A.S. Argon, A stochastic model for continuum elasto-plastic behavior. I. Numerical approach and strain localization, Model. Simul. Mater. Sci. Eng. 2 (1994) 167-184. 		

\bibitem{Bulatov1994b}
V.V. Bulatov and A.S. Argon, A stochastic model for continuum elasto-plastic behavior. II. A study of the glass transition and structural relaxation, Model. Simul. Mater. Sci. Eng. 2 (1994) 185-202. 			

\bibitem{Bulatov1994c}
V.V. Bulatov and A.S. Argon, A stochastic model for continuum elasto-plastic behavior. III. Plasticity in ordered versus disordered solids, Model. Simul. Mater. Sci. Eng. 2 (1994) 203-222. 

\bibitem{Homer2009}
E.R. Homer and C.A. Schuh, Mesoscale modeling of amorphous metals by shear transformation zone dynamics, Acta Mater. 57 (2009) 2823-2833. 			

\bibitem{Homer2010}
E.R. Homer and C.A. Schuh,  Three-dimensional shear transformation zone dynamics model for amorphous metals, Model. Simul. Mater. Sci. Eng. 18 (2010) 065009. 

\bibitem{Li2013}
L. Li, E.R. Homer and C.A. Schuh, Shear transformation zone dynamics model for metallic glasses incorporating free volume as a state variable, Acta Mater. 61 (2013) 3347-3359.		

\bibitem{RodneyPRL2009}
D. Rodney, C. Schuh. Distribution of Thermally Activated Plastic Events in a Flowing Glass, Phys. Rev. Lett. 102 (2009) 235503.			

\bibitem{RodneyPRB2009}
D. Rodney, C.A. Schuh. Yield stress in metallic glasses: The jamming-unjamming transition studied through Monte Carlo simulations based on the activation-relaxation technique, Phys. Rev. B 80 (2009) 184203-184212.			

\bibitem{Koziatek2013}
P. Koziatek, J.-L. Barrat, P. M. Derlet, D. Rodney. Inverse Meyer-Neldel behavior for activated processes in model glasses, Phys. Rev. B 87 (2013) 224105.	

\bibitem{Swayamjyoti2014}
S. Swayamjyoti, J.F. L\"{o}ffler, and P.M. Derlet. Local structural excitations in model glasses, Phys. Rev. B 89 (2014) 224201.		

\bibitem{Fan2017}
Y. Fan, T. Iwashita, and T. Egami, Energy landscape-driven non-equilibrium evolution of inherent structure in disordered material, Nat. Comm. 8 (2017) 15417.

\bibitem{Swayamjyoti2016}
S. Swayamjyoti, J.F. L\"{o}ffler, and P.M. Derlet, Local structural excitations in model glass systems under applied load, Phys. Rev. B 93 (2016) 144202.	

\bibitem{Derlet2017}
P.M. Derlet, R. Maass. Thermal processing and enthalpy storage of an amorphous solid: a molecular dynamics study, J. Mat. Res. 32 (2017) 2668.		

\bibitem{Wahnstrom1991}
G. Wahnstr\"{o}m, Molecular-dynamics study of a supercooled two-component Lennard-Jones system, Phys. Rev. A 44 (1991) 3752-3764.				

\bibitem{Zemp2014}
J. Zemp, M. Celino, B Sch\"{o}nfeld, and J. F L\"{o}ffler. Icosahedral superclusters in Cu$_{64}$Zr$_{36}$ metallic glass. Phys. Rev. B. 90 (2014) 144108.	

\bibitem{Zemp2016}
J. Zemp, M. Celino, B. Sch\"{o}nfeld, and J. F. L\"{o}ffler. Crystal-like rearrangements of icosahedra in simulated copper-zirconium metallic glasses and their effect on mechanical properties, Phys. Rev. Lett. 115 (2015) 165501.		

\bibitem{Pedersen2010}
U. R. Pedersen, T. B. Schr{\o}der, J. C. Dyre, and P. Harrowell, Geometry of Slow Structural Fluctuations in a Supercooled Binary Alloy, Phys. Rev. Lett 104 (2010) 105701.		

\bibitem{Rodney2011}
D. Rodney , A. Tanguy , D. Vandembroucq , Modeling the mechanics of amorphous solids at different length scale and time scale, Model. Simul. Mater. Sci. Eng. 19 (2011) 083001.

\bibitem{Daw1984}
M.S. Daw, M. Baskes, Embedded-atom method: derivation and application to impurities, surfaces, and other defects in metals, Phys. Rev. B 29 (1984) 6443-6453.

\bibitem{Finnis1984}
M. W. Finnis and J. E. Sinclair, A simple empirical N-body potential for transition metals, Phil. Mag. A 50 (1984) 45-55 

\bibitem{Rycroft2009}
C. H. Rycroft, Voro++: A three-dimensional Voronoi cell library in C++, Chaos 19 (2009) 041111.		

\bibitem{Stukowski2010}
A. Stukowski, Visualization and analysis of atomistic simulation data	with OVITO–the Open Visualization Tool, Modelling Simul. Mater. Sci. Eng. 18 (2010) 015012.	

\bibitem{Ryltsev2016}
R. R. Ryltsev, B. A. Klumov, N. M. Chtchelkatchev, K. Y. Shunyaev, Cooling rate dependence of simulated Cu$_{64.5}$Zr$_{35.5}$ metallic glass structure, J. Chem. Phys. 145 (2016) 034506.	

\bibitem{Donati1998}
C. Donati, J. F. Douglas, W. Kob, S. J. Plimpton,P. H. Poole, and S. C. Glotzer, Phys. Rev. Lett. 80 (1998) 2338-2341.

\bibitem{Schroeder2000}
T. B. Schr\o{}der, S. Sastry, J. C. Dyre, and S. C. Glotzer, J. Chem. Phys. 112 (2000) 9834-9849.

\bibitem{Gebremichael2004}
Y. Gebremichael, M. Vogel, and S. C. Glotzer, J. Chem. Phys. 120 (2004) 4415-4427.

\bibitem{Vogel2004}
M. Vogel, B. Doliwa, A. Heuer, and S. C. Glotzer, J. Chem.Phys. 120 (2004) 4404-4414.

\bibitem{Chandler2010}
D. Chandler and J. P. Garrahan, Dynamics on the way to forming glass: bubbles in Space-Time, Annu. Rev. Phys. Chem. 61 (2010) 191-217.

\bibitem{Kawasaki2013}
T. Kawasaki and A. Onuki, J. Chem. Phys. 138 (2013) 12A514-9.

\bibitem{Shi2005}
Y. Shi, M.L. Falk. Strain Localization and Percolation of Stable Structure in Amorphous Solids, Phys. Rev. Lett. 95 (2005) 095502.			

\bibitem{Chaudhuri2007}
P. Chaudhuri, L. Berthier and W. Kob,Universal Nature of Particle Displacements close to Glass and Jamming Transitions, Phys. Rev. Lett. 99 (2007) 060604.	

\bibitem{Ciamarra2016}
M. P. Ciamarra, R. Pastore and A. Coniglio, Particle jumps in structural glasses, Soft Matter 12 (2016) 358-366.			

\bibitem{Frank1958}
F. C. Frank and J. S. Kasper, Complex Alloy Structures Regarded as Sphere Packings. I. Definitions and Basic Principles, Acta Crystallogr. 11 (1958) 184.

\bibitem{Eshelby1957}
J.D. Eshelby, The Determination of the Elastic Field of an Ellipsoidal Inclusion, and Related Problems, Proc. Roy. Soc. A 241 (1957) 376-396.

\bibitem{Baumer2013}
R. E. Baumer and M.J. Demkowicz, Glass Transition by Gelation in a Phase Separating Binary Alloy, Phys. Rev. Lett 110 (2013) 145502.

\bibitem{Zhang2017}
M. Zhang, Y. M. Wang, F. X. Li, S. Q. Jiang, M. Z. Li and L. Liu , Mechanical Relaxation-to-Rejuvenation Transition in a Zr-based Bulk Metallic Glass, Sci. Rep. 7 (2017) 625.

\bibitem{Battezzati1984}
L. Battezzati, G. Riontino, M. Baricco, A. Lucci, F. Marino, A DSC study of structural relaxation in metallic glasses prepared with different quenching rates, J. Non-Cryst. Solids 61\&62 (1984) 877-882.
\end{thebibliography}
\end{document}